\documentclass[8.5pt,twoside,twocolumn]{article}
\usepackage[super,sort&compress,comma]{natbib} 
\usepackage{mhchem}
\usepackage{times,mathptmx}
\usepackage{sectsty}
\usepackage{balance} 

\usepackage{graphicx} 
\usepackage{lastpage}
\usepackage[format=plain,justification=raggedright,singlelinecheck=false,font=small,labelfont=bf,labelsep=space]{caption} 
\usepackage{fancyhdr}
\usepackage{multirow}

\usepackage{epsfig}
\usepackage{bm}        
\usepackage{amsmath} 
\usepackage{amsthm} 
\usepackage{amssymb}	
\usepackage{graphics} 
\usepackage{hyperref} 
\hypersetup{
    colorlinks,
    citecolor=blue,
    filecolor=black,
    linkcolor=blue,
    urlcolor=blue
}
\usepackage{epstopdf}

\begin{document}

\thispagestyle{plain}
\fancypagestyle{plain}{
\renewcommand{\headrulewidth}{1pt}}
\renewcommand{\thefootnote}{\fnsymbol{footnote}}
\renewcommand\footnoterule{\vspace*{1pt}%
\hrule width 3.4in height 0.4pt \vspace*{5pt}}
\setcounter{secnumdepth}{5}

\makeatletter 
\def\subsubsection{\@startsection{subsubsection}{3}{10pt}{-1.25ex plus -1ex minus -.1ex}{0ex plus 0ex}{\normalsize\bf}} 
\def\paragraph{\@startsection{paragraph}{4}{10pt}{-1.25ex plus -1ex minus -.1ex}{0ex plus 0ex}{\normalsize\textit}} 
\renewcommand\@biblabel[1]{#1}            
\renewcommand\@makefntext[1]%
{\noindent\makebox[0pt][r]{\@thefnmark\,}#1}
\makeatother 
\renewcommand{\figurename}{\small{Fig.}~}
\sectionfont{\large}
\subsectionfont{\normalsize} 

\fancyfoot{}
\fancyhead{}
\renewcommand{\headrulewidth}{1pt} 
\renewcommand{\footrulewidth}{1pt}
\setlength{\arrayrulewidth}{1pt}
\setlength{\columnsep}{6.5mm}
\setlength\bibsep{1pt}

\twocolumn[
  \begin{@twocolumnfalse}
\noindent\LARGE{\textbf{Stability of jammed packings II: the transverse length scale}}
\vspace{0.6cm}

\noindent\large{\textbf{Samuel S. Schoenholz,\textit{$^{a}$} Carl P. Goodrich,$^{\ast}$\textit{$^{a}$} Oleg Kogan,\textit{$^{a,b}$} Andrea J. Liu,\textit{$^{a}$} and
Sidney R. Nagel\textit{$^{b}$}}}\vspace{0.5cm}


\vspace{0.6cm}

\noindent \normalsize{As a function of packing fraction at zero temperature and applied stress, an amorphous packing of spheres exhibits a jamming transition where the system is sensitive to boundary conditions even in the thermodynamic limit.  Upon further compression, the system should become insensitive to boundary conditions provided it is sufficiently large.  Here we explore the linear response to a large class of boundary perturbations in 2 and 3 dimensions. We consider each finite packing with periodic-boundary conditions as the basis of an infinite square or cubic lattice and study properties of vibrational modes at arbitrary wave vector. We find that the stability of such modes be understood in terms of a competition between plane waves and the anomalous vibrational modes associated with the jamming transition; infinitesimal boundary perturbations become irrelevant for systems that are larger than a length scale that characterizes the transverse excitations.  This previously identified length diverges at the jamming transition.}
\vspace{0.5cm}
 \end{@twocolumnfalse}
  ]
\footnotetext{\textit{$^{a}$~Department of Physics, University of Pennsylvania, Philadelphia, Pennsylvania 19104, USA; E-mail: cpgoodri@sas.upenn.edu}}
\footnotetext{\textit{$^{b}$~James Franck and Enrico Fermi Institutes and Department of Physics, The University of Chicago, Chicago, Illinois 60637, USA}}

\section{Introduction}
At the jamming transition of ideal spheres, the removal of a single contact causes the rigid system to become mechanically unstable.~\cite{Epitome,ARCMP,FiniteSize}  Thus at the transition, the replacement of periodic-boundary conditions with free-boundary conditions destroys rigidity even in the thermodynamic limit.~\cite{Wyart2005,MatthieuLeo2005,Lstar} Recognizing that packings at the jamming threshold are susceptible to boundary conditions, Torquato and Stillinger~\cite{StrictJamming} drew a distinction between collectively jammed packings, which are stable when the confining box is not allowed to deform, and strictly jammed packings, which are stable to arbitrary perturbations of the boundary.  Indeed, Dagois-Bohy {\it et al.}~\cite{ShearInstability} have shown that jammed packings with periodic-boundary conditions can be linearly unstable to shearing the box.  

At densities greater than the jamming transition there are more contacts than the minimum required for stability.~\cite{Epitome,ARCMP} In this regime one would expect packings that are sufficiently large to be stable to changes in the boundary.  How does the characteristic size for a stable system depend on proximity to the jamming transition?  
The sensitivity to free- versus periodic-boundary conditions is governed by a length scale, $\ell^*$, that diverges at jamming transition. For $L \gg \ell^*$, the system is stable even with free boundaries.~\cite{Wyart2005,MatthieuLeo2005,Lstar}

Our aim in this paper is to identify the range of system pressures and sizes over which the system is likely to be unstable to a more general class of boundary perturbations in which particle displacements violate periodic boundary conditions.  We will show that stability is governed by a competition between transverse plane waves and the anomalous modes that are unique to disordered systems. Thus, we show that stability for a large class of boundary perturbations is governed by a 
length scale, $\ell_\text{T}$, that also diverges at the jamming transition.  Packings with $L \gg \ell_\text{T}$ are linearly stable with respect to these boundary perturbations. We understand this as a competition between jamming transition physics at low pressures/system sizes, and transverse acoustic wave physics at high pressures/system sizes. The two lengths, $\ell^*$ and $\ell_\text{T}$, are the same as the longitudinal and transverse lengths associated with the normal modes of jammed sphere packings.~\cite{TransverseLength} We will discuss the physical meaning of these length scales in more detail in sec.~\ref{sec:two_lengths}.

We analyze athermal, frictionless packings with periodic boundary conditions composed of equal numbers of small and large spheres with diameter ratio 1:1.4 all with equal mass, $m$. The particles interact via the repulsive finite-range harmonic pair potential
\begin{equation}\label{eq:potential}V(r_{ij}) = \frac{{\varepsilon}}2 \left(1-\frac{r_{ij}}{\sigma_{ij}}\right)^2 \end{equation}
if $r_{ij}<\sigma_{ij}$ and $V(r_{ij}) = 0$ otherwise. Here $r_{ij}$ is the distance between particles $i$ and $j$, $\sigma_{ij}$ is the sum of the particles' radii, and $\varepsilon$ determines the strength of the potential. Energies are measured in units of $\varepsilon$, distances in units of the average particle diameter $\sigma$, and frequencies in units of $\sqrt{{\varepsilon}/m \sigma^2}$.  We varied the total number of particles from $N= 32$ to $N=512$ at 36 pressures between $p=10^{-1}$ and $p=10^{-8}$. Particles are initially placed at random in an infinite temperature, $T=\infty$, configuration and are then quenched to a $T=0$ inherent structure using a combination of linesearch methods, Newton's method and the FIRE algorithm.\cite{FIRE} The resulting packing is then compressed or expanded uniformly in small increments until a target pressure, $p$, is attained.  After each increment of $p$, the system is again quenched to $T=0$.

\section{Symmetry-breaking perturbations}
\begin{figure}[h!tb]
\centering
\includegraphics[width=0.45\textwidth]{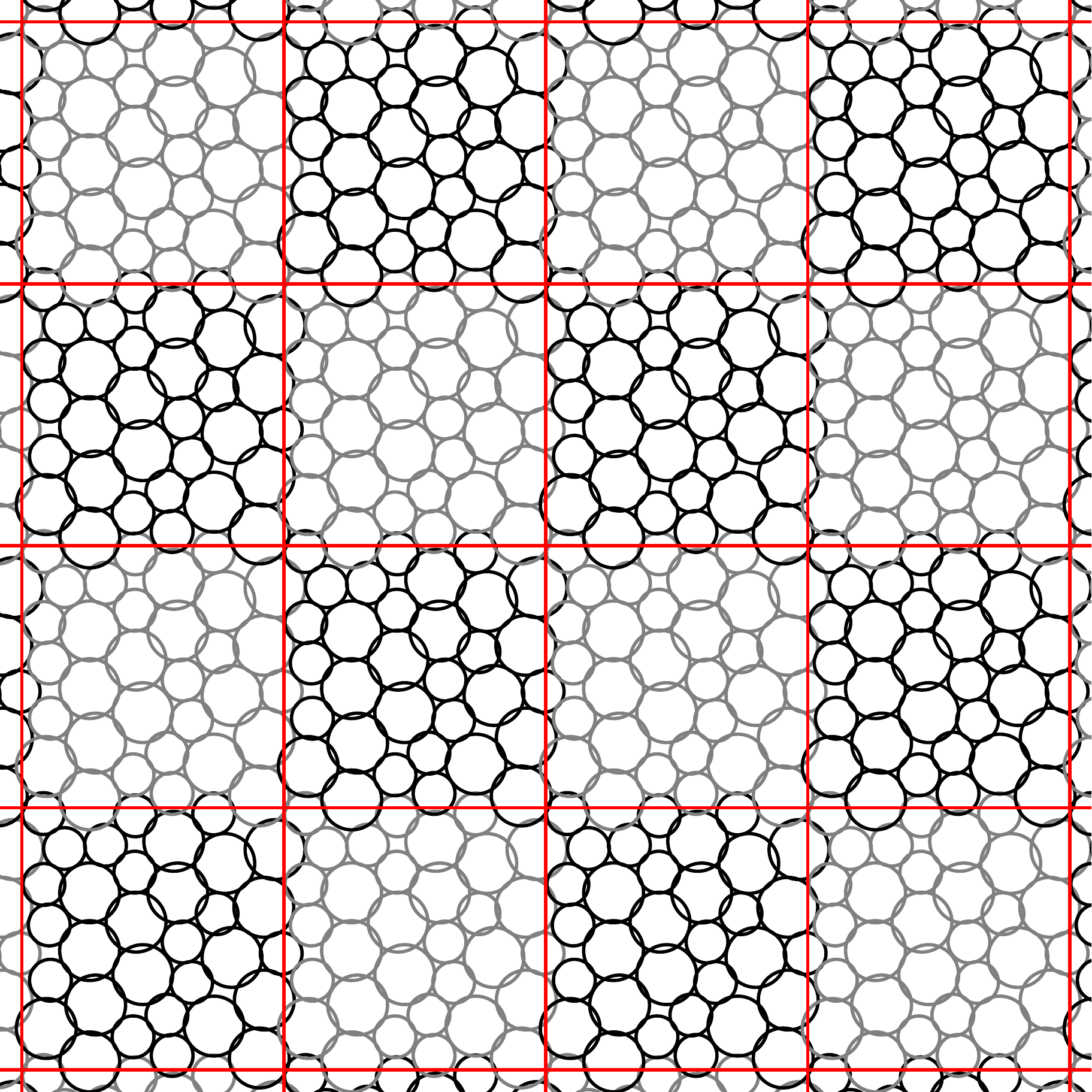}
\caption{A 32 particle system 
with periodic boundary conditions interpreted as a tiling in space. Here a 4$\times$4 section of the tiling is shown. Shading is used to provide contrast between adjacent copies of the system.}
\label{fig:checker}
\end{figure}

The boundary conditions can be perturbed in a number of ways. The dramatic change from periodic to free boundaries has been studied in refs.~\cite{Wyart2005,MatthieuLeo2005,Lstar}. Dagois-Bohy {\it et al.}~\cite{ShearInstability} considered infinitesimal ``shear-type" deformations to the \emph{shape} of the periodic box, such as uniaxial compression, shear, dilation, etc. Here, we relax the periodic boundary conditions by considering a third class of perturbations that allow particle displacements that violate periodicity. To do this, we treat our system with periodic boundary conditions as a tiling of identical copies of the system over all space (see Fig.~\ref{fig:checker}). Thus, an $N$-particle packing in a box of linear size $L$ with periodic boundary conditions can be viewed as the $N$-particle unit cell of an infinite hypercubic lattice.

\begin{figure}[h!tb]
a: Stable dispersion \qquad c: Example stable mode\\
\includegraphics[width=0.5\textwidth]{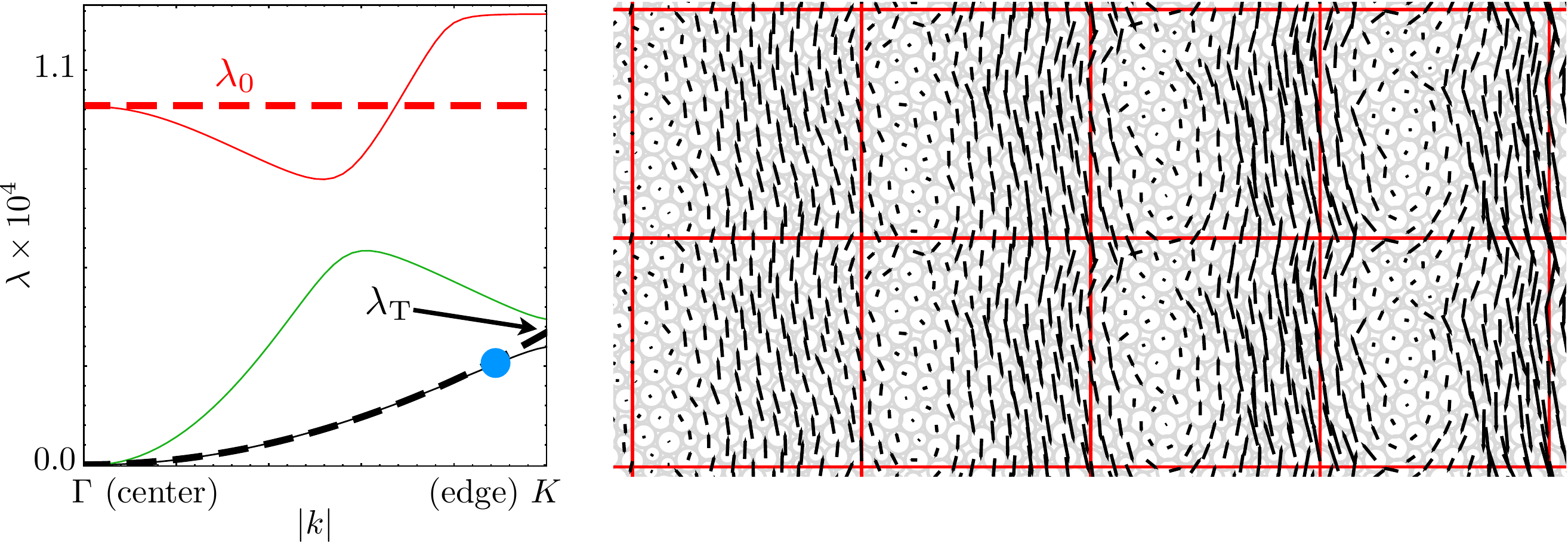}\\
b: Unstable dispersion \quad d: Example unstable mode \\
\includegraphics[width=0.5\textwidth]{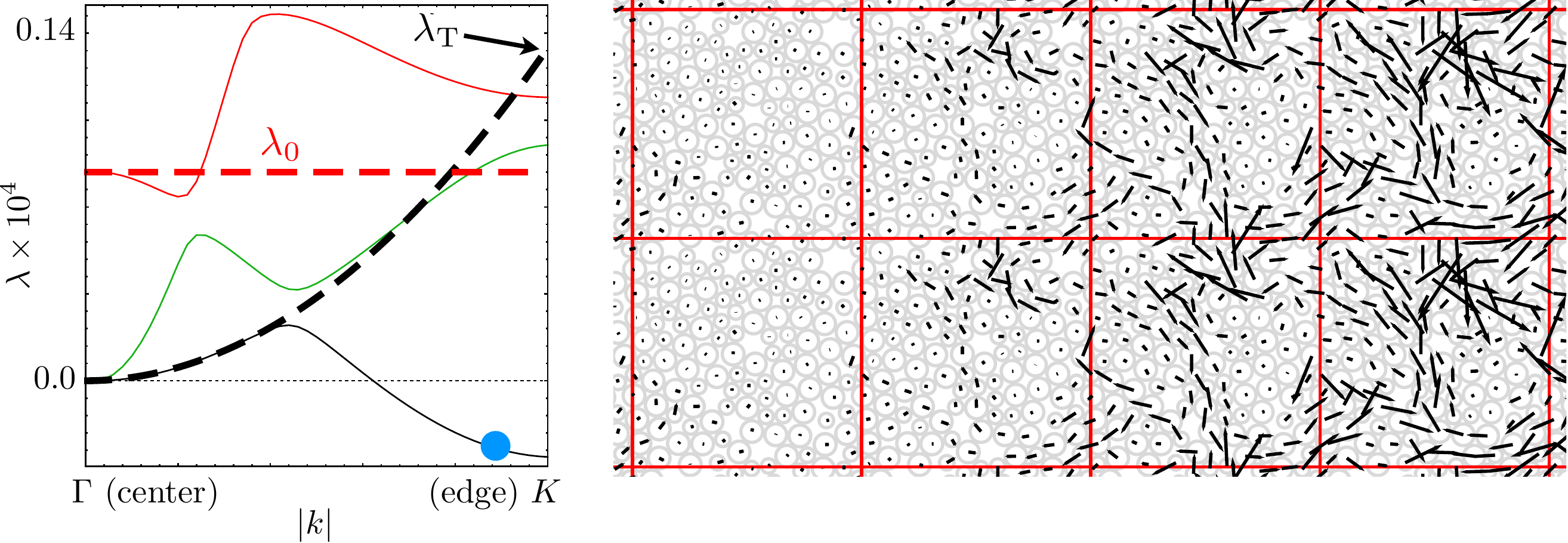}
\caption{(a) The dispersion relation for the transverse-acoustic mode (black), the longitudinal-acoustic mode (green), and the lowest anomalous mode (red) along the horizontal axis for a system of 128 particles in two dimensions at a pressure of $p=10^{-1}$. The black dashed line shows the quadratic approximation to the transverse branch and has a value of $\lambda_\text{T}$ at the zone edge. The dashed horizontal red line shows the flat approximation of the lowest anomalous mode and has a value of $\lambda_0$. The system is stable because the lowest branch is never negative. (b) Similar dispersion relations for a system at $p=10^{-3}$. This system has a lattice instability because the lowest mode is negative at large $k$. (c)-(d) A $2\times 4$ section of the periodically tiled systems from (a)-(b), respectively. Overlaid are the displacement vectors $\bm u_{i\mu}$ for the lowest modes near the zone edge ($\bm k\approx0.9\pi/L\bm {\hat x}$, see the blue dot in (a)-(b)). Note that the mode in (c) has strong transverse plane-wave character, while the mode in (d), which is unstable because $\lambda<0$, has developed strong anomalous character.}
\label{fig:pw-anom}
\end{figure}

The normal modes of vibration can be found by solving the equations of motion to linear order.  To do this, we assume the particles begin in mechanical equilibrium at positions specified by $\bm r_{i\mu}^0$, where $i$ indexes particles in each cell and $\mu$ indexes unit cells, so that $\bm r_{i\mu}^0$ is the equilibrium position of particle $i$ in cell $\mu$. 
The energy of the system to lowest order in particle displacements about its minimum value, $\bm u_{i\mu} = \bm r_{i\mu} - \bm r_{i\mu}^0$, can generically be written as
\begin{equation}\label{eq:linearization}U = \sum_{{\left<i\mu,j\nu\right>}}V(r_{i\mu j\nu}) \sim U_0 + \sum_{{\left<i\mu,j\nu\right>}}\bm u_{i\mu}^\text{T} \frac{\partial^2 U}{\partial \bm r_{i\mu}\partial \bm r_{j\nu}}\bigg|_{\bm r= \bm r^0}\bm u_{j\nu},
\end{equation}
where the sums are over all pairs of particles $i\mu$ and $j\nu$ that are in contact.
The equations of motion resulting from eqn~(\ref{eq:linearization}) can be solved by a plane-wave ansatz,
$\bm u_{i\mu} = \mathrm{Re}\left\{\bm \epsilon_i \exp\left[i(\bm k\cdot \bm R_\mu-\omega t)\right]\right\}$.
Here $\bm \epsilon_i$ is a $dN$-dimensional polarization vector, $\bm k$ is a $d$-dimensional wavevector and $\bm R_\mu$ is the $d$-dimensional vector corresponding to the position of cell $\mu$. This gives the eigenvalue equation
\begin{equation}\label{eq:eigenvalue}\lambda_n(\bm k)\bm \epsilon_{ni}(\bm k) = \sum_{j}\bm D_{i j}(\bm k)\bm \epsilon_{nj}(\bm k),\end{equation}
where
\begin{equation}\label{eq:dynamical}\bm D_{ij}(\bm k) = \sum_{\mu\nu}\frac{\partial^2  U}{\partial \bm r_{i\mu}\partial \bm r_{j\nu}}\bigg|_{\bm r= \bm r^0}e^{i\bm k\cdot(\bm R_\mu - \bm R_\nu)}\end{equation}
is the dynamical matrix of dimension $dN \times dN$, and $n$ labels the eigenvalues and eigenvectors. From eqn~(\ref{eq:linearization}), the frequency of the modes in the $n$th branch are $\omega_n(\bm k)=\sqrt{\lambda_n(\bm k)}$ with eigenvector $\bm \epsilon_{ni}(\bm k)$. Fig.~\ref{fig:pw-anom} shows $\lambda_{n}(\bm k)$ as a function of $\bm k$ for two example systems, as well as as well as examples of $\bm u_{i\mu}$ that solve the eigenvalue equation. The behavior of these examples will be discussed in detail below.

With $\bm k$ allowed to vary over the first Brillouin zone, the eigenvectors comprise a complete set of states for the entire tiled system. It follows that any displacement of particles at the boundary of the unit cell can be written as a Fourier series,
\begin{equation}\label{eq:fourierseries}\bm u_{i\mu} = \sum_{\bm k,n}A_{\bm k,n}\bm \epsilon_{ni}(\bm k) \exp\left[i(\bm k\cdot \bm R_\mu)\right].\end{equation}
Therefore, the system will be unstable to some collective perturbation of its boundary if and only if there is some $\bm k$ and $n$ for which $\lambda_n(\bm k)< 0$. This procedure allows us to characterize boundary perturbations by wavevector.\footnote[3]{Note that shear-type deformations can be considered concurrently by adding the term $\Lambda \cdot \bm r_{i\mu}^0$ to eqn~\eqref{eq:fourierseries}, where $\Lambda$ is a global deformation tensor. This term represents the affine displacement and is neglected for our purposes.} If we find a wavevector whose dynamical matrix has a negative eigenvalue, it follows that the system must be unstable with respect to the boundary perturbation implied by the corresponding eigenvector. We will show that stability is governed by a competition between transverse plane waves and anomalous modes, examples of which are shown in Fig.~\ref{fig:pw-anom}.

The gist of our argument can be understood as follows.  The lowest branch of the normal mode spectrum is composed of transverse plane waves; it is these modes that are the ones most easily perturbed by disorder to produce a negative eigenfrequency and therefore a lattice instability.  The largest perturbation comes from an interaction with the higher-frequency anomalous modes which have a characteristic frequency of $\omega^*$.  If $\omega^*$ is large, which would occur at high pressure, or if the highest-frequency transverse mode ({\it i.e.}, at the zone boundary) is small, which would occur for large systems, then there is unlikely to be a strong perturbation and all the modes will remain positive and there will be no instability.  

To illustrate, we show in Fig.~\ref{fig:pw-anom}a the lowest three eigenvalue branches along the horizontal axis of the Brillouin zone for an example system that is shown in Fig.~\ref{fig:pw-anom}c. At low $k$, the lowest two branches correspond to transverse and longitudinal plane waves, respectively, and increase quadratically in $k$. The third branch corresponds to the lowest anomalous mode in the system. When the longitudinal branch approaches the anomalous mode, band repulsion pushes the branch down to lower energies. However, the transverse branch maintains its quadratic behavior all the way to the zone edge. Near the zone edge, modes of this lowest branch have very strong plane-wave character, as shown by the displacement vectors in Fig.~\ref{fig:pw-anom}c.

Contrast this to the dispersion curves shown in Fig.~\ref{fig:pw-anom}b, 
for a system at much lower pressure. Here, the lowest anomalous branch is roughly an order of magnitude lower than in the previous example, and therefore has a more drastic effect on the lowest two branches. Specifically, the transverse branch experiences band repulsion about half way to the zone edge, and the quadratic extrapolation (the dashed black line) is a very poor predictor of the eigenvalue at the edge. In this second example, modes near the edge have negative eigenvalues, meaning they are unstable. Furthermore, as illustrated in Fig.~\ref{fig:pw-anom}d, these unstable modes do not resemble plane-waves, but instead appear disordered and closely resemble anomalous modes. 

This scenario is generic. 
We define $\lambda_\text{T}$ to be the energy of the transverse branch extrapolated to the zone boundary and $\lambda_0$ to be the energy of the lowest anomalous branch (see Fig.~\ref{fig:pw-anom}). When $\lambda_0$ is much larger than $\lambda_\text{T}$, then the lowest mode maintains its transverse plane-wave character throughout the Brillouin zone. However, when $\lambda_0$ is less than $\lambda_\text{T}$, then the lowest mode at the zone edge is anomalous and is more likely to be unstable. Importantly, plane waves and anomalous modes are governed by different physics. By understanding the scaling of $\lambda_0$ and $\lambda_\text{T}$, we will show that stability is controlled by the transverse length scale, $\ell_\text{T}$.

Our strategy will be to first consider the so-called ``unstressed system," which replaces the original system of spheres with an identical configuration of particles connected by unstretched springs with stiffness given by the original bonds. In the unstressed system, eigenvalues can never be negative and so all systems are stable.  We will then construct scaling relations for $\lambda_0$ and $\lambda_{\text T}$ of the unstressed system and conclude that the lowest eigenvalue will be $\min\{\lambda_0,\lambda_{\text T}\}$. We then construct additional scaling relations for the accompanying shift in the eigenvalues upon reintroducing the stress. By finding the pressures and system sizes where these two quantities are comparable, we will thus determine the scaling of the susceptibility of packings to perturbations described by eqn~\eqref{eq:fourierseries}.

\section{The unstressed system}
The dynamical matrix in eqn~(\ref{eq:dynamical}) is a function of the second derivative of the pair potentials of eqn~\eqref{eq:potential} with respect to particle positions. To construct the dynamical matrix of the unstressed system, we rewrite this as as
\begin{align}
\label{eq:stressed}
\frac{\partial^2V}{\partial r_{\alpha}\partial r_{\beta}} &= \frac{\partial^2 V}{\partial r^2}\frac{\partial r}{\partial r_{\alpha}}\frac{\partial r}{\partial r_{\beta}} + \frac{\partial V}{\partial r}\frac{\partial^2 r}{\partial r_{\alpha}\partial r_{\beta}},
\end{align}
where $V$ is the potential between particles $i\mu$ and $j\nu$, $\bm r \equiv \bm r_{j\nu} - \bm r_{i\mu}$ and $\alpha$ and $\beta$ are spatial indices. 
The second term is proportional to the negative of the force between particles. If this term is neglected, there are no repulsive forces and the system will be ``unstressed."~\cite{MatthieuLeo2005,Alexander199865}
The dynamical matrix of the unstressed system, obtained from {\it just} the first term in eqn~\eqref{eq:stressed}, has only non-negative eigenvalues 
. We will use a subscript ``$\text{u}$" (as in, {\it e.g.}, $\lambda_{\text{u}}$) to refer to quantities corresponding to the unstressed system.

\begin{figure*}
\centering
\includegraphics[width=1.0\textwidth,viewport=110 10 1180 345,clip]{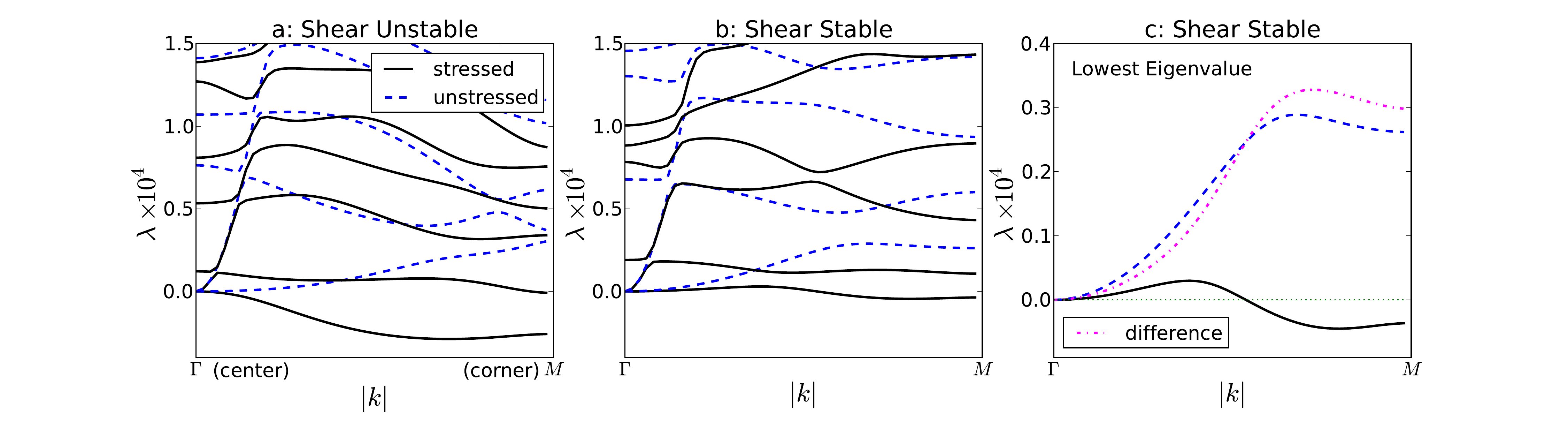}
\caption{Dispersion relations along the $\Gamma-M$ line for the lowest few branches of 2 different 2-dimensional packings of $N=1024$ particles.  The $\Gamma$ point is at the Brillouin zone center ($\bm k = \bm 0$) and the $M$ point is at the zone corner where the magnitude of $\bm k$ is greatest. (a)  A ``shear-unstable'' packing (\emph{i.e.} a sphere packing that is unstable at low $k$) at $p = 10^{-4}$ (black) and the dispersion relation for the corresponding unstressed system (blue). (b) A ``lattice-unstable'' sphere packing (\emph{i.e.} a packing that is stable near $k=0$ but is unstable at higher $k$) at $p=10^{-4}$ (black) and the dispersion relation for the corresponding unstressed system (blue). (c) Comparison between the lowest branch in the sphere packing (black) and its unstressed counterpart (blue) for the same system as in (b).  The dashed magenta line is the difference between the two eigenvalue branches.}
\label{fig:examplemodes}
\end{figure*}

The black curves in Fig.~\ref{fig:examplemodes}a show the six lowest eigenvalue branches, $\lambda({\bm k})$, for a 2-dimensional packing of $N=1024$ disks at a pressure $p = 10^{-4}$.
Also shown (dashed blue curves) are the lowest eigenvalue branches for the corresponding unstressed system, $\lambda_\text{u}({\bm k})$, where the second term in eqn~\eqref{eq:stressed} has been omitted.  Here, the lowest branch of the stressed system has negative curvature at $k=0$ implying that the system has a negative shear modulus.  Such shear-type instabilities have been analyzed previously.~\cite{ShearInstability}  In the remainder of this paper, we restrict our attention to shear-stable packings that are stable near $k=0$ but potentially unstable at higher wave vectors. We refer to this type of instability as a ``lattice instability."

Fig.~\ref{fig:examplemodes}b compares the 6 lowest eigenvalue branches of a sphere packing (black) with a lattice instability to those of its unstressed counterpart (dashed blue).  The lowest branch for the sphere packing has positive curvature at $k=0$, but becomes negative at higher $k$.  This implies that the system is unstable to boundary perturbations corresponding to eqn~\eqref{eq:fourierseries} over a range of wavevectors. In contrast, the unstressed system, by construction, can have only positive (or possibly zero) eigenvalues.   In Fig.~\ref{fig:examplemodes}c, the dotted magenta line shows the difference between the lowest eigenvalue branch of the unstressed system (blue) and of the sphere packing (black).

\section{The lowest eigenvalue of the unstressed system}  
We first evaluate the eigenvalues in the unstressed system and then consider the effect of the stress term (\emph{i.e.} the second term in eqn~\eqref{eq:stressed}). In order to obtain the scaling of the lattice instabilities, we first estimate the eigenvalues at the $M$ point, which is located at the corner of the Brillouin zone. The wave vector $\bf k_M$ at the corner has the smallest wavelength anywhere in the Brillouin zone and thus corresponds to the most drastic perturbation. 
We then extend the argument to the rest of the Brillouin zone. 

For the unstressed system, the lowest eigenvalue at the corner of the Brillouin zone can be estimated as follows.  As Fig.~\ref{fig:examplemodes}b suggests, the mode structure is fairly straightforward. 
Low-frequency vibrations are dominated by two distinct classes of modes: plane waves and the so-called ``anomalous modes" that are characteristic of jammed systems.~\cite{Wyart2005,ARCMP}
The lowest plane-wave branch is transverse and parabolic at low $k$ (see Fig.~\ref{fig:examplemodes}b): $\omega_{\text{T},\text{u}} \approx c_{\text{T},\text{u}}k$ or equivalently
\begin{equation}
\lambda_{\text{T},\text{u}} \approx c_{\text{T},\text{u}}^2 k^2 \sim G_\text{u} k^2, \label{eq:parabola}
\end{equation}
where $G_\text{u}$ is the shear modulus of the unstressed system.

The eigenvalue of the lowest anomalous mode can be understood as follows. Wyart {\it et al.}~\cite{Wyart2005,MatthieuLeo2005} showed that the density of vibrational states at $k=0$ for unstressed systems, $D_\text{u}(\omega)$, can be approximated by a step function, so that $D_\text{u}(\omega) \approx 0$ for $\omega<\omega_\text{u}^*$ while $D_\text{u}(\omega) \approx \mathrm{const}$ for $\omega>\omega_\text{u}^*$.  As suggested by Fig.~\ref{fig:examplemodes}, the anomalous modes are nearly independent of $\bm k$, so this form for $D_\text{u}(\omega)$ is a reasonable approximation not only at $k=0$ but over the entire Brillouin zone. 
Thus, the eigenvalue of the lowest anomalous mode is approximated by $\omega_\text{u}^*$ at any $\bm k$.

Note that if $\omega_{\text{T},\text{u}} \ll \omega_\text{u}^*$, the lowest branch will maintain its transverse-acoustic-wave character and hence will remain parabolic in $k$ all the way to the zone corner.  
However, when $\omega_{\text{T},\text{u}} \gg \omega_\text{u}^*$, the lowest mode at the corner no longer has plane-wave character because the transverse acoustic mode will hybridize with anomalous modes and will develop the character of those modes.  It follows that there is a crossover between jamming physics and plane-wave physics when $\omega_{\text{T},\text{u}} \approx \omega_\text{u}^*$ or, equivalently, when the system size is
\begin{equation}\label{eq:crossoverlength}L \approx \ell_\text{T}\equiv c_{\text{T},\text{u}}/\omega_\text{u}^*.\end{equation}
Here $\ell_\text{T}$ is the transverse length identified by Silbert {\it et al.}~\cite{TransverseLength}

Near the jamming transition, many properties scale as power laws with the excess contact number above isostaticity, $\Delta Z \equiv Z - Z^N_\mathrm{iso}$, where $Z$ is the average number of contacts per particle and $Z^N_\mathrm{iso} = 2d \left(1-1/N\right) \approx 2d$ for large systems.~\cite{FiniteSize} In particular, for the harmonic potentials we consider here, $\omega_\text{u}^*\sim \Delta Z$,  $G_\text{u} \sim \Delta Z$, and $p\sim\Delta Z^2$ for dimensions $d \ge 2$. (These results are easily generalized to potentials other than the harmonic interactions used here.~\cite{ARCMP})
Therefore, eqns~\eqref{eq:parabola} and \eqref{eq:crossoverlength} predict that, for $d \ge 2$, $\ell_\text{T} \sim p^{-1/4}$ and the crossover will occur at 
\begin{equation}\label{eq:crossover} pL^4 \sim\text{const.}\end{equation}

\begin{figure}[!h]
\centering
\includegraphics[width=0.49\textwidth,viewport=0 75 500 650,clip]{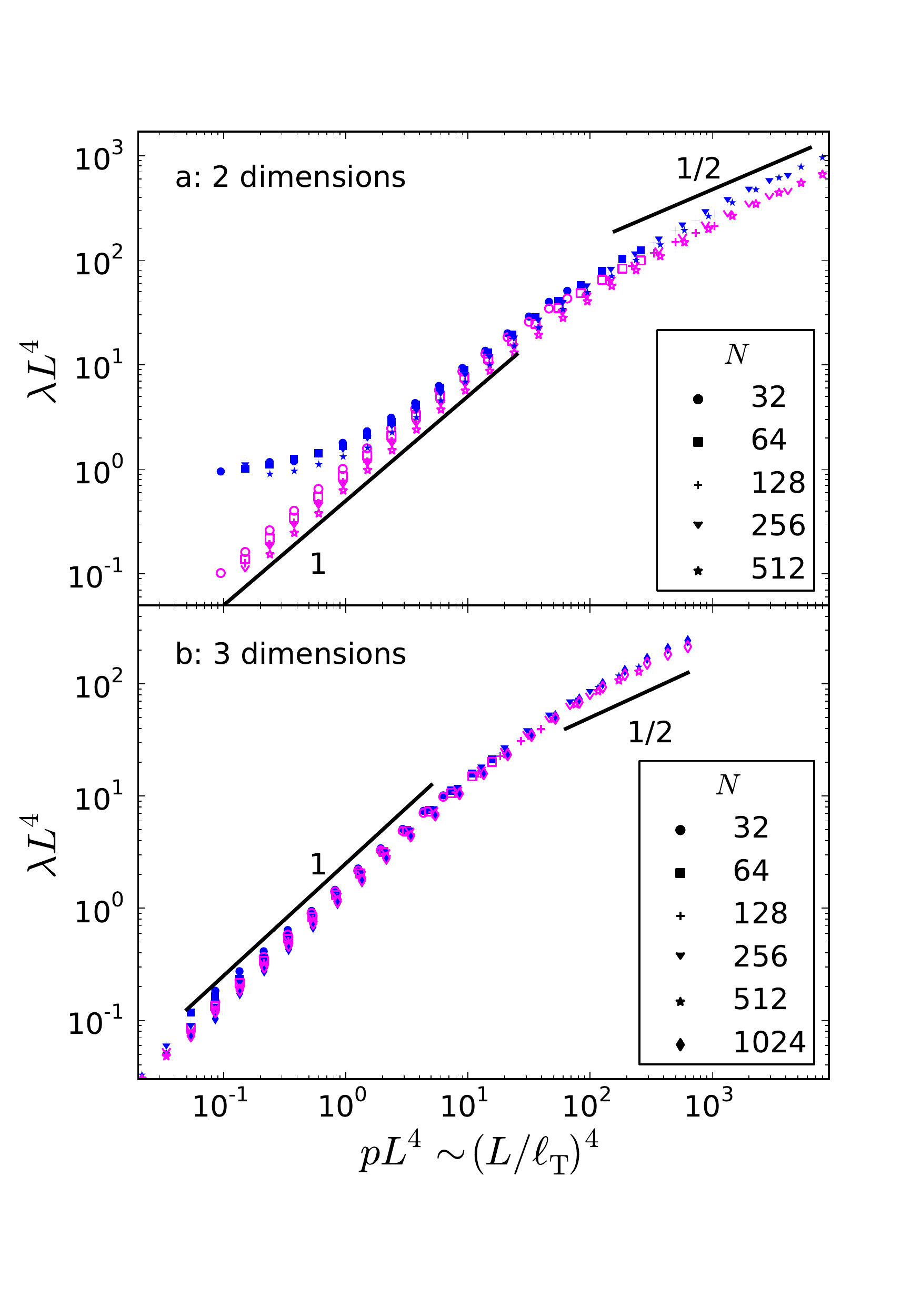}
\caption{The average eigenvalue of the lowest mode at the corner of the Brillouin zone of the unstressed system (blue), as well as the average difference between the unstressed system and the original packing (magenta), in (a) two and (b) three dimensions. The blue and magenta data exhibit collapse as predicted by eqns~\eqref{eq:lambda0} and \eqref{eq:stressdiff}, with the caveat that we were unable to reach the low pressure regime in three dimensions where we expect the scaling to be different. Data is only shown when at least 20 shear stable configurations were obtained.}
\label{fig:stressedunstressed}
\end{figure} 

A second crossover occurs at very low pressures and is due to finite-size effects~\cite{FiniteSize} that change the scaling of $\Delta Z$ to $\Delta Z\sim1/N\sim L^{-d}$ in $d$ dimensions, independent of $p$.  In this regime, $\omega_\text{u}^*$ and $G_\text{u}$ remain proportional to $\Delta Z$, and thus also scale as $L^{-d}$. We therefore expect the lowest eigenvalue of the unstressed system at the corner of the Brillouin zone ($k_{M} = \sqrt{d}\pi/L$) to feature three distinct regimes.
\begin{equation}
	\begin{array}{ r r l}
		\text{low pressure:} &  		\lambda_\text{u} & \sim \;\;  1/N^2 \sim L^{-2d} \\
		\text{intermediate pressure:} &	\lambda_\text{u}  & \sim \;\; \omega^{*2}_\text{u} \sim p  \\
 		\text{high pressure:} &		\lambda_\text{u}  & \sim \;\; c_\text{T}^2 k_M^2 \sim p^{1/2} L^{-2}. \label{eq:lambda0}
	\end{array}
\end{equation}
In two dimensions, we expect that $\lambda_\text{u}L^4$ will collapse in all three regimes as a function of $pL^4$.

This prediction is verified in Fig.~\ref{fig:stressedunstressed}. The blue symbols in Fig.~\ref{fig:stressedunstressed}a, corresponding to $\lambda_\text{u} L^4$ of the unstressed system in $d=2$, exhibit a plateau at low pressures/system sizes. At intermediate $pL^4$, the blue symbols have a slope of 1 and at high $pL^4$ a slope of 1/2, as predicted by eqn~\eqref{eq:lambda0}. 
In three dimensions (Fig.~\ref{fig:stressedunstressed}b), we observe the two higher pressure regimes, with a crossover between them that scales with $pL^{4}$, as expected. We did not reach the low-pressure plateau regime in three dimensions because it is difficult to generate shear stable configurations at low pressures.  Note, however, we do not expect the crossover to the low pressure regime to collapse in $d=3$ with $pL^4$.  

\begin{figure*}[htpb]
\centering
\includegraphics[width=0.95\textwidth,viewport=0 75 950 650,clip]{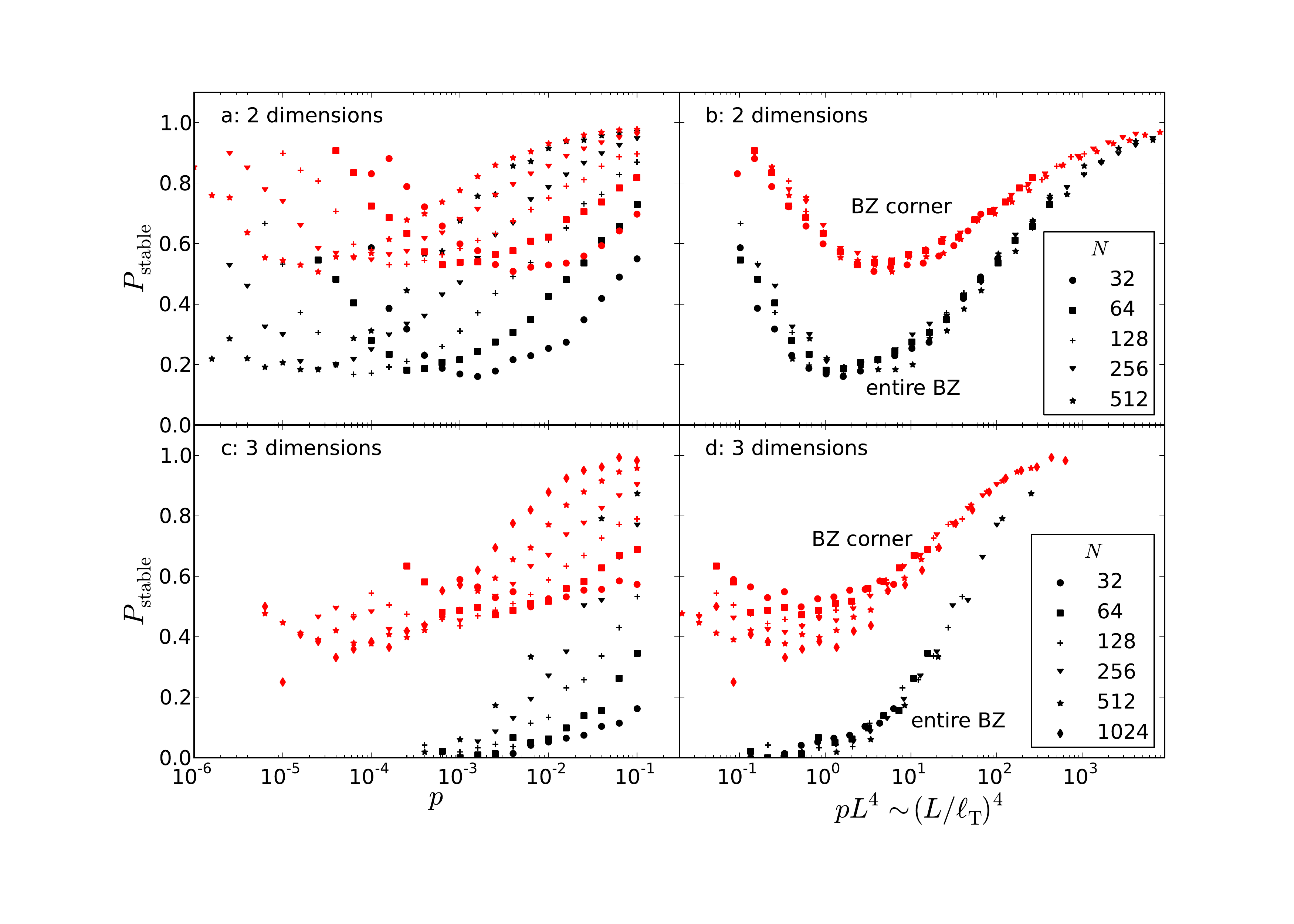}
\caption{The fraction of shear-stable systems in two and three dimensions that are also $k>0$-stable. In red we plot the fraction of systems that are unstable at the $M$ point while in black we plot the fraction of systems that are stable everywhere. We see that both collapse with $pL^4$, or equivalently $L/\ell_\text{T}$, with the expected exception of the low pressure regime in three dimensions.}
\label{fig:goodapples}
\end{figure*}

\section{The effect of stress on the lowest eigenvalue}
With the behavior of the lowest eigenvalue of the unstressed system in hand, we now turn to the effects of stress to explore the behavior of actual sphere packings at the zone corner, $k_{M} = \sqrt{d}\pi/L$.  The second term in eqn~\eqref{eq:stressed} shifts the shear modulus to smaller values without affecting the scaling with pressure~\cite{Ellenbroek:2009to}, $G \sim \sqrt{p}$.   It therefore follows that  $G_\text{u}-G\sim\sqrt p$. In the high-pressure regime, where the lowest mode is the transverse plane wave, we therefore expect that $\lambda_\text{u}-\lambda \sim G_\text{u}-G \sim \sqrt{p}$.   Likewise, it was shown~\cite{Lstar,WyartThesis,XuPRL2007} that the second term in eqn~\eqref{eq:stressed} lowers ${\omega_\text{u}^*}^2$ by an amount proportional to $p$, so at intermediate and low pressures it follows that $\lambda_\text{u}-\lambda \approx{\omega_\text{u}^*}^2-{\omega^*}^2 \sim p$.

The difference in the lowest eigenvalues of the sphere packing and the unstressed system will therefore feature two distinct regimes: 
\begin{equation}
	\begin{array}{ r r l}
		\text{low and int. $p$:} &	\left(\lambda_\text{u}-\lambda\right) & \sim p  \\
 		\text{high $p$:} &		\left(\lambda_\text{u}-\lambda\right) & \sim p^{1/2} L^{-2}. \label{eq:stressdiff}
	\end{array}
\end{equation}
 
\section{The lowest eigenvalue and stability of the original packing}
Comparing these results to eqn~\eqref{eq:lambda0}, we see that the lowest eigenvalue of the packing $\lambda$ should be positive in the low pressure limit where $\lambda_\text{u}-\lambda \ll \lambda_\text{u}$.  At high and medium pressures, $\lambda_\text{u}$ and $\lambda_\text{u}-\lambda$ are comparable and obey the same scaling. One would therefore expect instabilities to arise in this regime.  At high pressures, however, we know that $\lambda$ should be positive since $\lambda \sim G k_M^2$ and shear stability implies $G>0$.~\cite{ShearInstability}   Therefore, fluctuations about the average scaling behavior are most likely to drive the system unstable at intermediate pressures.
Since this regime collapses with pressure and system size as $pL^4$, we expect the fraction of systems that are unstable to obey this scaling.  This prediction is corroborated in Fig.~\ref{fig:goodapples}, where we see that the fraction of systems that are stable at the $M$ point (red data) depends on pressure $p$ and system size $L$ as expected.

We have thus far illustrated our reasoning for wavevectors at a zone corner, which correspond to the smallest wavelengths and thus the most drastic perturbations. Systems should therefore be more likely to go unstable at the zone corner than anywhere else. However, our arguments apply equally well to any point in the Brillouin zone. 

This argument is confirmed in Fig.~\ref{fig:goodapples}. To investigate the stability over the entire zone, we computed the dispersion relation over a fine mesh in $k$ space and looked for negative eigenvalues.  Any configuration with a negative eigenvalue at any value of $\bm k$ was labeled as unstable.  The black data in Fig.~\ref{fig:goodapples} shows the fraction of systems that are stable over the entire Brillouin zone. It exhibits the same qualitative features as at just the zone corner (red data), but with fewer stable configurations overall. 

Note that our scaling arguments for the unstressed system apply equally well to the original packing. However, scaling alone cannot tell us the {\it sign} of the lowest eigenvalue. Treating the unstressed and stressed terms of the dynamical matrix separately, and exploiting the fact that the unstressed matrix is non-negative, is thus necessary for understanding the stability of the packing and emphasizes that ``lattice instabilities" are caused by stress, not the geometry of the packing.

\section{The two length scales\label{sec:two_lengths}}
Dagois-Bohy {\it et al.}~\cite{ShearInstability} found that in two dimensions, the fraction of states that are shear stable collapses with $pL^4$ and approaches $1$ at large $pL^4$. 
We have shown that shear stable packings can be unstable to a class of boundary perturbations that correspond to particle motion at non-zero $\bm k$, and that the susceptibility of packings to such perturbations also collapses with $pL^4$ (except for the $N$-dependent finite-size effects at low pressure). 
The combination of these two results suggests that a system should be stable to all infinitesimal boundary perturbations for $L \gg 
\ell_\text{T}$, where $\ell_\text{T}\sim p^{-1/4}$ (for harmonic interactions) is the so-called transverse length scale.  This implies that the closer the system is to the jamming transition the larger the system must be in order to be insensitive to changes in the boundaries.  It also implies that the distinction between collectively and strictly jammed~\cite{StrictJamming} packings decreases as $L \gg \ell_\text{T}$ and vanishes in the thermodynamic limit for any nonzero pressure.

While the sensitivity of jammed packings to infinitesimal changes to the boundary is controlled by the diverging length scale $\ell_\text{T}$, Wyart {\it el al.}~\cite{Wyart2005,MatthieuLeo2005} argued that the stability to a more drastic change of boundary conditions, in which the periodic boundaries are replaced with free ones, is governed by the larger length $\ell^*\sim p^{-1/2}$: jammed packings with \emph{free} boundaries are stable only for $L \gg \ell^*$.~\cite{Wyart2005,MatthieuLeo2005,Lstar} 
Goodrich {\it et al.}~\cite{Lstar} have shown that this length is equivalent, not only in scaling behavior but also in physical meaning, to the longitudinal length $\ell_\text{L} =\omega^*/c_\text{L}$. 

$\ell_\text{L}$ was proposed alongside $\ell_\text{T}$ by Silbert {\it et al.},~\cite{TransverseLength} and both length scales can be understood from a competition between plane-wave physics and jamming physics. Silbert {\it et al.} showed that at large wavelengths, there are well defined transverse and longitudinal sound modes; the Fourier transform $f_\text{T,L}(\omega,k)$ is sharp in $\omega$ at low $k$.~\cite{TransverseLength,Silbert:2009iw} However, at small wavelengths, the sound modes mix with the anomalous modes and $f_\text{T,L}$ gets very broad. The crossover between these two regimes occurs when the frequency of the sound mode is $\omega^*$, the lowest frequency at which anomalous modes exist. Since the frequency of the sound modes is given by $\omega = c_\text{T,L}k$, the wave length at the crossover is $\ell_\text{T,L} \sim c_\text{T,L}/\omega^*$.
Thus, the two length scales $\ell_\text{T}$ and $\ell^*=\ell_\text{L}$ are both related to the wavelength where the density of anomalous modes overwhelms the density of plane-wave sound modes.~\cite{TransverseLength} Our results indicate that they also both control the stability of jammed packings to different classes of boundary perturbations. The two lengths can therefore be viewed as equally central to the theory of the jamming transition.

\section{Discussion}
To understand why some boundary perturbations are controlled by $\ell_\text{T}$ while others are controlled by $\ell_\text{L}$, we must understand the nature of the transverse and longitudinal deformations allowed.  In the case of systems with free boundaries, the system size of $\ell_\text{L}$ needed to support compression is also sufficient to support shear since $\ell_\text{L}>\ell_\text{T}$.  The minimum system size needed to support both is therefore $\ell_\text{L}$. Under periodic boundary conditions, jammed systems necessarily support compression regardless of system size and the issue is whether the system is stable to shear as well.  Our arguments show that the minimum system size needed to support both longitudinal and transverse perturbations is $\ell_\text{T}$.  

The same reasoning can be applied to understanding the stability with respect to a change from periodic to fixed boundary conditions.  In that case, neither longitudinal nor transverse deformations are allowed and it follows that jammed systems of any size will be stable to this change. Mailman and Chakraborty~\cite{Mailman:2011hz} have studied systems with fixed boundary conditions to calculate point-to-set correlations. Their analysis, as well as arguments based on the entropy of mechanically stable packings, reveal a correlation length that scales like $\ell^*$.

The fact that the diverging length scales control the response to boundary changes but do not enter into the finite-size scaling of quantities such as the contact number and shear modulus~\cite{FiniteSize} is consistent with the behavior of a system that is at or above its upper critical dimension.  The fact that power-law exponents do not depend on dimensionality for $d \ge 2$ is also consistent with this interpretation.~\cite{Epitome,zamponi}  However, critical systems are generally controlled by only one diverging length.  Jamming is thus a rare example of a phase transition that displays two equally important diverging length scales. 
It remains to be seen whether other diverging lengths that have been reported near the jamming transition~\cite{Epitome,Reichhardts,Teitelshear,Teiteldistbns,Mailman:2011hz} can also be associated with boundary effects.

We thank David L. Johnson for prompting us to study vibrations of jammed tilings, Justin Burton for suggestions regarding energy minimization methods, as well as Wouter Ellenbroek, Tom Lubensky, and Anton Souslov for stimulating discussions.  This research was supported by the U.S. Department of Energy, Office of Basic Energy Sciences, Division of Materials Sciences and Engineering under Award DE-FG02-05ER46199 (AJL,SS,OK) and DE-FG02-03ER46088 (SN,OK), and by the NSF via the graduate research fellowship program (CPG) and the UPenn MRSEC DMR11-20901 (SS).


\balance



\footnotesize{
\providecommand*{\mcitethebibliography}{\thebibliography}
\csname @ifundefined\endcsname{endmcitethebibliography}
{\let\endmcitethebibliography\endthebibliography}{}

}

\end{document}